\documentclass[prd,twocolumn,floatfix,nofootinbib]{revtex4}
\usepackage{amssymb}
\usepackage{amsmath}
\usepackage{graphicx,subfigure,color,dcolumn,booktabs,bm}
\usepackage{longtable,lscape}
\usepackage{txfonts}
\usepackage{overpic}
\usepackage{indentfirst}
\usepackage{cases}
\usepackage{multirow}
\usepackage{ulem}
\usepackage[colorlinks,
            citecolor=blue,
            anchorcolor=red,
            menucolor=red,
            linkcolor=red,
            filecolor=red,
            runcolor=red,
            urlcolor=blue,
            frenchlinks=false]{hyperref}

\graphicspath{{Figures/}}
\allowdisplaybreaks
\begin{document}

\title{Masses and decays of triply-heavy pentaquarks}
\author{Chang-Le Liu $^{1}$}
\email{liuchanglelcl@qq.com}
\author{Wen-Xuan Zhang $^{1}$}
\email{zhangwx89@outlook.com}
\author{Duojie Jia$^{2,3}$}
\email{jiadj@nwnu.edu.cn, Corresponding author}
\affiliation{$^1$Institute of Theoretical Physics, College of Physics and Electronic
Engineering, Northwest Normal University, Lanzhou 730070, China \\
$^2$General Education Center, Qinghai Institute of Technology, Xining, 810000, China \\
$^3$Lanzhou Center for Theoretical Physics, Lanzhou University, Lanzhou,
730000, China \\
}
\date{\today}

\begin{abstract}

In this work, we study masses and decays of triply-heavy pentaquarks $QQQn\bar{n}(Q=b,c)$ in the unified MIT bag model. We construct the color-spin wave functions of the triply-heavy pentaquarks we address and use numerical variational method to compute all ground-state masses of these system.  By excluding the scattering states in these configurations, we compute the decay width ratios of each decay channels relative to the maximum width for the compact pentaquark states, obtaining main decay modes of the triply-heavy pentaquark systems.

PACS number(s):12.39Jh, 12.40.Yx, 12.40.Nn

\end{abstract}

\maketitle
\date{\today}

\section{Introduction}
\label{sec:intro}

In 1964, independently, Gell-Mann\cite{gell214} and Zweig\cite{Zweig:1964jf, Zweig:1964ruk} formulated a quark model using quarks as fundamental constituents. This model provide a more systematic understanding of numerous hadronic systems, suggesting the possible existence of multiquark states beyond the conventional hadrons. In the 1970s, several theories and models emerged to explore these multiquark states\cite{Richard:2012xw}, including the MIT bag model\cite{Chodos:1974pn,Yan:2023lvm}, constituent quark model\cite{Ortega:2023uhh}, string model\cite{Richard:2009rp,Ryskin:2006wk}, QCD sum rules\cite{Wang:2023jaw,Wang:2023kir,Albuquerque:2023gke} and their variants, as well as lattice QCD methods\cite{Bicudo:2022cqi}.

In comparison to conventional hadronic states, our understanding of exotic hadronic states remains relatively limited. After years of experimental searches, in 2003, the Belle experiment discovered a distinct particle, the X(3872)\cite{PhysRevLett.91.262001}, which markedly differed from traditional hadronic states. In 2015, the LHCb experiment observed structures resembling pentaquark states, namely $P_{c}(4380)^{+}$ and $P_{c}(4450)^{+}$, in the decay $\Lambda^{0}_{b} \to J/ \psi pK^{-}$\cite{PhysRevLett.115.072001}. In 2019, the $P_{c}(4450)^{+}$ state was resolved into a two-peak structure comprising $P_c(4440)^{+}$ and $P_{c}(4457)^{+}$\cite{LHCb:2019kea}. In 2021, the LHCb collaboration discovered two tetraquark states, $Z_{cs}(4000)^{+}$ and $Z_{cs}(4220)^{+}$, containing a strange quark in the decay process $B^{+}\to J/\psi\phi K^{+}$\cite{PhysRevLett.127.082001}. Furthermore, observations such as $P_{cs}$\cite{PhysRevLett.131.031901,LHCb:2020jpq}, $Z_{c}(3900)^{+}$\cite{PhysRevLett.110.252001,PhysRevLett.110.252002}, $Z_{c}(4430)^{+}$\cite{PhysRevLett.100.142001,PhysRevD.88.074026}, and others have significantly advanced research related to multiquark states, particularly regarding their mass spectra and decay behaviors. Since then, several peculiar hadronic states or candidates have been discovered in high-energy physics experiments, making the investigation of exotic hadronic states one of the focal points.

\begin{figure}
\centering
\includegraphics[width=0.45\textwidth]{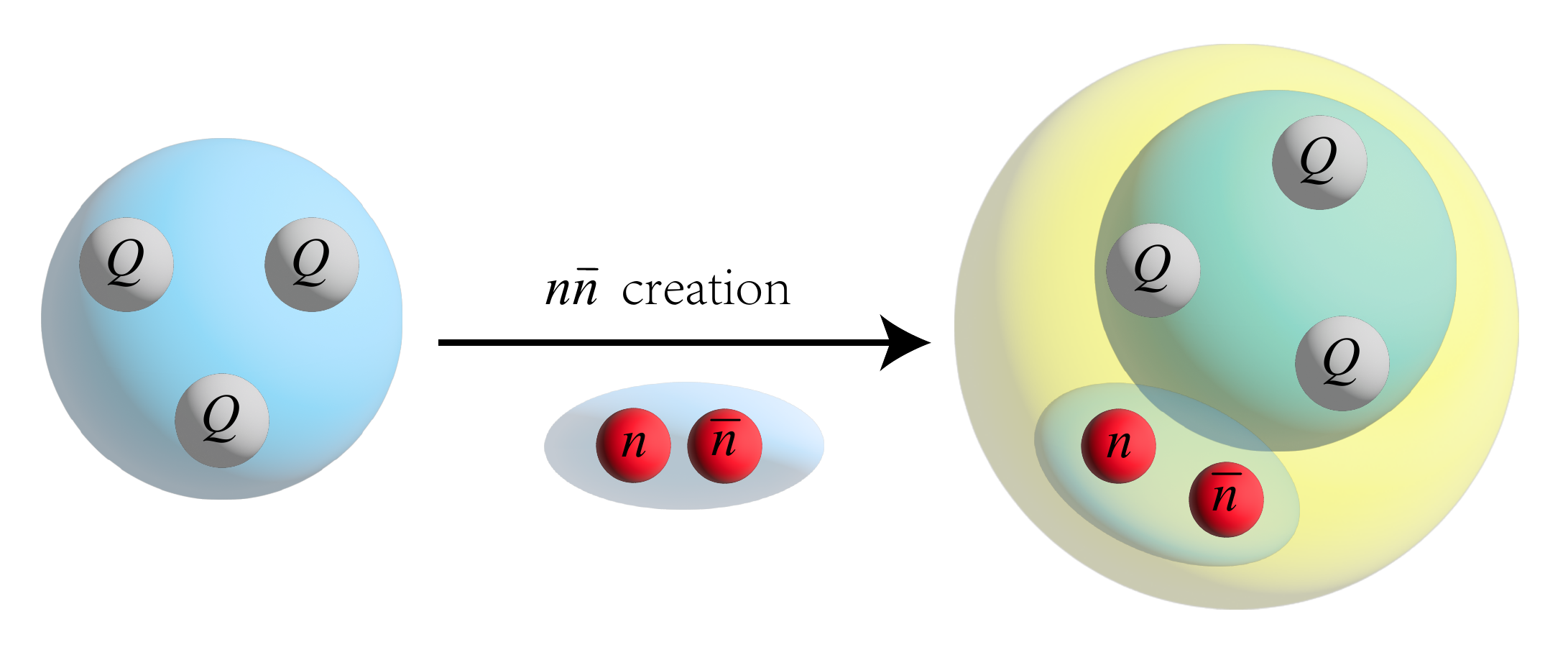}
\caption{The triply-heavy baryon $QQQ$ combines with the creation of a light-flavored meson $n\bar{n}$ to form the triply-heavy pentaquark state $QQQn\bar{n}$.}
\label{fig:figa}
\end{figure}

Since the experimental confirmation of the doubly charmed baryon $\Xi_{cc}^{++}$\cite{LHCb:2017iph,SELEX:2002wqn}, theoretical investigations have commenced on multiquark states containing two or more heavy quarks\cite{Zhou:2018bkn,An:2019idk,Zhang:2023hmg,Cheng:2022vgy,PhysRevD.105.034006,G:2024zkc,Wang:2023eng}. Studies have explored potential triply-charm molecular pentaquarks such as $\Xi_{cc}D_{1}$ and $\Xi_{cc}D_{2}^{\ast}$\cite{Wang:2019aoc}, and compact pentaquark states with a $qqQQ\bar{Q}$ configuration ( where $q=n, s$; $Q=c, b$ )\cite{An:2019idk}. Motivated by these findings, our research focuses on the configuration of the triply-heavy pentaquark state $QQQn\bar{n}$. In this configuration, it can be perceived as a pentaquark state formed after the creation of $n\bar{n}$, combined with a $QQQ$, As shown in Fig.(\ref{fig:figa}).

The aim of this study is to investigate the masses and partial decay width ratios of the triply-heavy pentaquark state $QQQn\bar{n}$, categorized based on the different heavy flavor quarks into $cccn\bar{n}$, $ccbn\bar{n}$, $bbcn\bar{n}$, and $bbbn\bar{n}$. The MIT bag model has broad applications in baryons\cite{He:2004px}, mesons\cite{Zhang:2021yul}, pentaquark states, and hybrids\cite{BARNES1983241,CHANOWITZ1983211}. In the framework of MIT bag model, the masses of $QQQn\bar{n}$ are calculated by using specific model parameters and fundamental relations.  In addition, based on the partial decay width ratios, we further discuss the decay modes of various specific configurations.

The structure of this work is organized as fellows: In Section \uppercase\expandafter{\romannumeral2}, we introduce the MIT bag model, which is utilized to calculate the theoretical masses of $QQQn\bar{n}$. Section \uppercase\expandafter{\romannumeral3} outlines the methodology for computing partial decay width ratios and contains four subsections presenting the calculated results and decay mode analyses for $cccn\bar{n}$, $ccbn\bar{n}$, $bbcn\bar{n}$, and $bbbn\bar{n}$, listing the decay products accordingly. Finally, in Section \uppercase\expandafter{\romannumeral4}, we ended with a summary.

\section{MIT bag model}
\label{sec:mothed}

The MIT bag model describes hadron as a fundamental physical representation of valence quarks confined within a spherical “bag". For a hadron described by a bag of radius $R$, the model provides a mass formula expressed as \cite{PhysRevD.12.2060,Johnson:1975pj}:

\begin{equation}
    \begin{aligned}
        M(R) =\sum_{i}^{}\omega_{i}+\frac{4}{3} \pi R^{3}B-\frac{Z_{0}}{R}+M_{BD}+M_{CMI},
    \end{aligned}
    \label{eq:eq1}
\end{equation}

\begin{equation}
    \begin{aligned}
        \omega_{i}=(m_{i}^{2}+\frac{x_{i}^{2}}{R^{2}})^{1/2}.\\
    \end{aligned}
    \label{eq:eq2}
\end{equation}

The first term in the equation denotes the cumulative relativistic kinetic energy of all valence quarks confined within the bag. For any quark $i$, the relativistic kinetic energy comprises both the mass $m_{i}$ and momentum $\frac{x_{i}}{R}$. The parameter $x_{i}$ is related to the bag radius $R$ by the equation below
\cite{PhysRevD.12.2060,Johnson:1975pj}:

\begin{equation}
    \begin{aligned}
          \tan x_{i}=\frac{x_{i}}{1-m_{i}R-(m_{i}^{2}R^{2}+x_{i}^{2})^{1/2}}.\\
    \end{aligned}
    \label{eq:eq3}
\end{equation}

This equation is obtained by applying bag surface boundary conditions to the following doublet spinor wave function within the context of the bag model:

\begin{equation}
\psi_{i}\left(r\right)=N_{i}\begin{pmatrix}
 j_{0}\left(x_{i}r/R\right)U \\
i\frac{x_{i}}{\left(\omega _{i}+m_{i}\right)R}j_{1}\left(x_{i}r/R\right)\sigma\cdot\hat{r}U
\end{pmatrix}e^{-i\omega_{i}t}.
\label{eq:eq4}
\end{equation}

The bag radius $R$ is determined using variational method, whereas $x_{i}$ is iteratively solved using the above Eq.(\ref{eq:eq3})\cite{PhysRevD.12.2060}.

The second term is the volume energy, where the constant $B$ denotes the energy density difference between perturbative and non-perturbative QCD vacuum. The presence of the third term accounts for zero-point energy, crucial for maintaining the bag's overall stability.

The final two terms in Eq.(\ref{eq:eq1}) represent the interaction between quarks. The  $M_{BD}$ term denotes the binding energy within the confinement bag between two heavy quarks or between a heavy quark and a strange quark\cite{PhysRevD.90.094007,Karliner:2017elp,Karliner:2018bms}. These binding energies can be expressed in a constant form\cite{Zhang:2021yul}:

\begin{equation}
\begin{Bmatrix}
  B_{cs}=-0.025\,\mathrm{GeV},& B_{cc}=-0.077\,\mathrm{GeV}, \\
  B_{bs}=-0.032\,\mathrm{GeV},& B_{bb}=-0.128\,\mathrm{GeV}, \\
  B_{bc}=-0.101\,\mathrm{GeV}.
\end{Bmatrix}
\end{equation}

The fifth term, $M_{CMI}$, is known as chromomagnetic interaction, representing the interaction between quarks confined within the bag by the lowest-order gluon exchange\cite{DeRujula:1975qlm}. The chromomagnetic interaction $M_{CMI}$ is expressed as follows\cite{Zhang:2021yul}:

\begin{equation}
    \begin{aligned}
       M_{CMI}=-\sum_{i<j}^{}(\lambda_{i}\cdot \lambda_{j})(\sigma_{i} \cdot \sigma_{j})C_{ij}.
    \end{aligned}
    \label{eq:eq6}
\end{equation}

In the context of the chromomagnetic interaction formula, the subscripts $i$ and $j$ represent indices for quarks or antiquarks, $\lambda$ stands for the Gell-Mann matrices, $\sigma$ denotes the Pauli matrices, and $C_{ij}$ represents the coupling parameters. For the color and spin factors within the chromomagnetic interaction formula, we employ the following matrix element formula:

\begin{equation}
     \begin{aligned}
    \left\langle\lambda_{i}\cdot\lambda_{j}\right\rangle_{nm}=
    \sum_{\alpha=1}^{8}\mathrm{Tr}(c_{in}^{\dagger}\lambda^{\alpha}c_{im})
    \mathrm{Tr}(c_{jn}^{\dagger}\lambda^{\alpha}c_{jm}),
    \end{aligned}
    \label{eq:eq7}
\end{equation}

\begin{equation}
     \begin{aligned}
    \left\langle\sigma_{i}\cdot\sigma_{j}\right\rangle_{xy}=\sum_{\alpha=1}^{3}
   \mathrm{Tr}(\chi _{ix}^{\dagger}\sigma ^{\alpha}\chi_{iy})\mathrm{Tr}(\chi _{jx}^{\dagger}\sigma^{\alpha}\chi _{jy}).
    \end{aligned}
    \label{eq:eq8}
\end{equation}

The subscript $n, m$ in Eq.(\ref{eq:eq7}) and $x, y$ in Eq.(\ref{eq:eq8}) represent the fundamental vector components of the color and spin wave functions for hadrons, with $c$ and $\chi$ symbolizing the color and spin vector bases relevant to quarks. Once the color-spin wave functions are established, the matrices for these two factors can be calculated using the above equations.

For the parameter $C_{ij}$ in the chromomagnetic interaction $M_{CMI}$, the following relation is satisfied\cite{PhysRevD.12.2060}:

\begin{equation}
    \begin{aligned}
        C_{ij}=3\frac{\alpha_{s}(R)}{R^{3}}\bar{\mu}_{i}\bar{\mu}_{j}I_{ij}.
    \end{aligned}
    \label{eq:eq9}
\end{equation}

In the mass formula Eq.(\ref{eq:eq1}) of the MIT bag model, all parameters except for $R$ and $x_{i}$ are constants. Parameters are represented as follows: $Z_{0}$, signifying the ground state energy constant; $B$, representing the bag constant; and $m_{i}$, indicating the mass of the quark species $i$ or its respective antiquark \cite{Zhang:2021yul}:

\begin{equation}
\begin{Bmatrix}
  Z_{0}=1.83,    & B^{1/4}=0.145\,\mathrm{GeV}, \\
  m_{n}=0\,\mathrm{GeV},    & m_{s}=0.279\,\mathrm{GeV}, \\
  m_{c}=1.641\,\mathrm{GeV},& m_{b}=5.093\,\mathrm{GeV}.
\end{Bmatrix}
\end{equation}

Since we do not consider the isospin effects (both of $m_{n=u,d}=0$), the masses for the isoscalar and isovector $QQQn\bar{n}$ pentaquarks are degenerate. When we mention a state, we mean all isospin multiplets of this state and assume the readers assign its isospin themselves.
By utilizing the given parameters, we limit the variables to only two components: $R$ and $x_{i}$. The parameter $x_{i}$ in momentum represents a solution to a transcendental equation, serving as an intermediary reliant on the variable $R$. Initially, an estimated value for $x_{i}$ is applied to Eq.(\ref{eq:eq3}) to solve for $R$.
One can then employ, for a given wave function composed of spatial part and color-spin part,  variational method to Eq.(\ref{eq:eq1}) and Eq.(\ref{eq:eq3}) to interatively solve  $R$ and $x_{i}$ consistently and thereby compute the masses of triply-heavy pentaquarks\cite{Zhang:2021yul,Zhang:2023hmg}.

\section{the decay channels of the $QQQn\bar{n}(Q=b,c)$ system}
\label{sec:result}

Using the MIT bag model from the previous section, the mass of the studied triply-heavy pentaquark state can be calculated. With the initial mass, one can further research the process of decay. Before diving into the specific study of the triply-heavy pentaquark state decay, certain scattering states need to be eliminated. To differentiate these scattering states from other the compact pentaquark states, it is necessary to employ the color-spin wave functions corresponding to each eigenvector of the triply-heavy pentaquark state.

The color-spin wave functions of the triply-heavy pentaquark state from the coupling of baryonic and mesonic decay products in two different ways: coupling between the baryon color singlet and meson color singlet, denoted as $1_{c}$, and coupling between the baryon color octet and meson color octet, denoted as $8_{c}$. The color wave functions corresponding to these two coupling modes for the triply-heavy pentaquark states are provided in the appendix \cite{Weng:2019ynv}.

\begin{equation}
\Psi=c_{1}\left|q_{1}q_{2}q_{3}\right\rangle^{1}_{S_{1}}
\left|q_{4}\bar{q}_{5}\right\rangle^{1}_{S_{2}}+
c_{2}\left|q_{1}q_{2}q_{3}\right\rangle^{8}_{S_{3}}
\left|q_{4}\bar{q}_{5}\right\rangle^{8}_{S_{4}}+\cdots
\label{eq:eq.11}
\end{equation}

For the color-spin wave function structure with the coefficient $1_{c}$ in the above equation, due to the coupling of S-wave baryon and meson via scattering state, they may produce a baryon with spin $S_{1}$ and a meson with spin $S_{2}$. If the pentaquark has a strong coupling with $1_{c}$, then the probability associated with this specific vector $\left |c_{1}\right |^{2}$ tends closer to 1. When the vector satisfies $\left |c_{1}\right |^{2}\ge0.8$, it will be identified as a scattering state. Such states are to be excluded. For the compact pentaquark state with the form $8_{c}$, it can also decay by exchanging quarks to convert $8_{c}$ into $1_{c}$.

When $QQQ\otimes n\bar{n}$ exchanges quarks, we can identify the compact states in $QQn\otimes Q\bar{n}$ configuration according to the compact pentaquark states found before. Because the symmetry of $QQn\otimes Q\bar{n}$ configuration is not as high as that of $QQQ\otimes n\bar{n}$ configuration, it will contain some non-physical states. Therefore, it is necessary to search for compact states under $QQQ\otimes n\bar{n}$ configuration. The decay channels of the compact pentaquark state can be further investigated once the scattering state is eliminated.

Here, we specifically study the two-body decay mode $A\to B+C$. For two-body decay, we can provide the partial width formula for each decay channels corresponding to the eigenvectors \cite{Weng:2019ynv,name2,Ruangyoo:2021aoi}:

\begin{equation}
\Gamma_{i}=\gamma_{i}\alpha\frac{k^{2L+1}}{m_{A}^{2L}}\cdot\left |c_{i}\right |^{2},
\label{eq:eq12}
\end{equation}

\begin{equation}
m_{A}=\sqrt{m_{B}^{2}+k^{2}}+\sqrt{m_{C}^{2}+k^{2}}.
\label{eq:eq13}
\end{equation}

In the above Eq.(\ref{eq:eq12}), $\Gamma_{i}$ denotes the partial width of decay channel $i$, while $\gamma_{i}$ represents a quantity determined by the dynamics of the decay process. $\alpha$ denotes the coupling constant, and $m_{A}$ corresponds to the mass of the initial compact pentaquark state before decay. The coefficient $c_{i}$ is the probability amplitude of the wave function calculated by diagonalization of chromomagnetic interaction matrix. $L$ represents orbital angular momentum, and since we specialize in ground states, let's set $L=0$. The symbol $k$ represents the momentum of the decay products  in the rest frame for the decay system. The momentum $k$ for the decay products can be computed using the Eq.(\ref{eq:eq13}) provided above. Additionally, $m_{B}$ and $m_{C}$ represent the masses of the baryon and meson produced in the decay, respectively. The mass parameters of the decay products are primarily sourced from the Particle Data Group \cite{Zhang:2021yul,PhysRevD.98.030001,SND:2020nwa}, while the mass parameters of the triply-heavy baryons are derived from calculations based on the MIT bag model \cite{Zhang:2023hmg}.

The decay coefficient $\gamma_{i}$ for the two-body decay $A\to B+C$ depends on the spatial wave functions of the initial and final states. For decay channels composed of scalar mesons (or vector mesons) and baryons with specific flavor combinations in the products, the corresponding $\gamma_{i}$ values are the same. For instance, if particle $A$ undergoes decay, and the resulting particle $B$ (a baryonic product) can be either $B(J=1/2)$ or $B^{\ast}(J=3/2)$, while the resulting particle $C$ (a mesonic product) can be either $C (J=0)$ or $C^{\ast} (J=1)$. For each decay channel in this decay process, the relation between the decay coefficients is satisfied as follows\cite{Weng:2019ynv,Weng:2021ngd,Weng:2022ohh}:

\begin{equation}
\gamma_{BC}=\gamma_{BC^{\ast}}=\gamma_{B^{\ast}C}=\gamma_{B^{\ast}C^{\ast}}.
\end{equation}

This relation applies to decay processes in which the baryons and mesons in the decay products possess a definite flavor configuration. We note here that for the degenerate pentaquarks of isoscalar and isovector, they have the same $QQn\otimes Q\bar{n}$ decay modes and relevant partial widths, indicating that the width ratios between the decays with final states having different isospin I are equal.  We also note that there are no isospin-dependent interactions in the adopted model for spectrum and width investigations.

For a compact pentaquark with a specific $J^{P}$ quantum number and mass, when the flavor compositions of the resulting baryons and mesons after decay are given, multiple decay channels can exist. Among these channels, we select the partial width of one decay channel as a reference standard. Then, we calculate the ratio of the partial widths of the remaining decay channels to that of the selected channel. This allows us to obtain the partial decay width ratios for all decay channels involving specific flavor combinations of baryons and mesons, along with their corresponding decay products.

For the triply-heavy pentaquark states, there exist three possible $J^{P}$ quantum numbers: $1/2^{-}$, $3/2^{-}$, and $5/2^{-}$. Here, we classify pentaquark states according to each $J^{P}$ quantum number and flavor configuration, and study them accordingly. The work on $QQQn\bar{n}(Q=b,c)$ in the following four subsections is mainly divided into two aspects:
on the one hand, the scattering states are distinguished from the compact pentaquark states; on the other hand, the decay channels and partial decay width ratios of the compact pentaquark states are provided.

\subsection{ $cccn\bar{n}$ system}

Using the MIT bag model approach and chromomagnetic interaction as described earlier, the masses and eigenvectors of the $cccn\bar{n}$ type pentaquark states are listed in Table \ref{table:tab1}. Let's first focus on the eigenvectors of the $cccn\bar{n}$ type pentaquark states. According to the color-spin wave functions in the appendix for the $QQQn\bar{n}$ type, we observe the following feature of the $J^{P}$ quantum numbers: for $J^{P}=1/2^{-}$, only the third coefficient in each eigenvector corresponds to a $1_{c}$ state; for $J^{P}=3/2^{-}$, both the second and third coefficients in the respective eigenvectors correspond to a $1_{c}$ state; and for $J^{P}=5/2^{-}$, there is only one coefficient in the eigenvector, which also corresponds to a $1_{c}$ state.

When the $J^{P}$ quantum number is $1/2^{-}$, the square of the coupling coefficient with $1_{c}$ in the eigenvector corresponding to the state with a mass of 5.741 GeV is below 0.8. Hence, this state cannot be defined as a scattering state. Similarly, the coefficients coupled to $1_{c}$ in 5.827 GeV and 5.963 GeV indicate that they cannot form a scattering state either. In summary, when $J^{P}=1/2^{-}$, the states of the $cccn\bar{n}$ configuration have no scattering states.

When $J^{P} = 3/2^{-}$, the square of the coefficients coupled to $1_{c}$ in the eigenvectors of 5.372 GeV and 5.786 GeV are greater than 0.8, indicating that both states are scattering states.

Particularly noteworthy is the case of $J^{P}=5/2^{-}$. Unlike color-spin wave function systems of the triply-heavy pentaquark states with $J^{P}$ quantum numbers of $1/2^{-}$ and $3/2^{-}$, the color-spin wave function system with $J^{P}=5/2^{-}$ in the $cccn\bar{n}$ configuration ($QQQn\bar{n}$ configuration in appendix) shows that only the scattering state exists. Therefore, the decay of the triply-heavy pentaquark state with respect to $J^{P}=5/2^{-}$ is no longer discussed in the $cccn\bar{n}$ configuration.

In Table \ref{table:tab1}, we list the decay products corresponding to the scattering states.
The scattering states in the blank represent the compact triply-heavy pentaquark states.

\renewcommand{\tabcolsep}{0.6cm}
\renewcommand{\arraystretch}{1.1}
\begin{table*}[!htbp]
    \caption{Masses, bag radius, eigenvectors, and scattering states of the triply-heavy pentaquark state $cccn\bar{n}$ system at each $J^{P}$ quantum number. The unit of mass is GeV and the unit of bag radius is $\mathrm {GeV^{-1}}$.}
    \begin{tabular}{c|ccccc}
        \hline\hline
        State &$J^{P}$ &$R_0$ &Mass &Eigenvector &Scattering state \\ \hline
        $cccn\bar{n}$
            &${1/2}^{-}$    &5.437  &5.741   &(0.302,  0.499,  0.812)  &$    $\\
            &$ $            &5.523  &5.827   &(-0.839, -0.273, 0.471)  &$    $\\
            &$ $            &5.665  &5.963   &(0.455,  -0.827, 0.330)  &$    $\\
            &${3/2}^{-}$    &5.424  &5.372   &(-0.108, 0, 0.994)       &$\Omega_{ccc}\pi$\\
            &$ $            &5.555  &5.786   &(0, 1, 0    )            &$\Omega_{ccc}\rho/\omega$\\
            &$ $            &5.546  &5.854   &(0.994,  0,   0.106 )    &$    $\\
            &${5/2}^{-}$    &5.555  &5.786            &(1)             &$\Omega_{ccc}\rho/\omega$\\
        \hline\hline
    \end{tabular}
    \label{table:tab1}
\end{table*}

\begin{figure}
\centering
\includegraphics[width=0.45\textwidth]{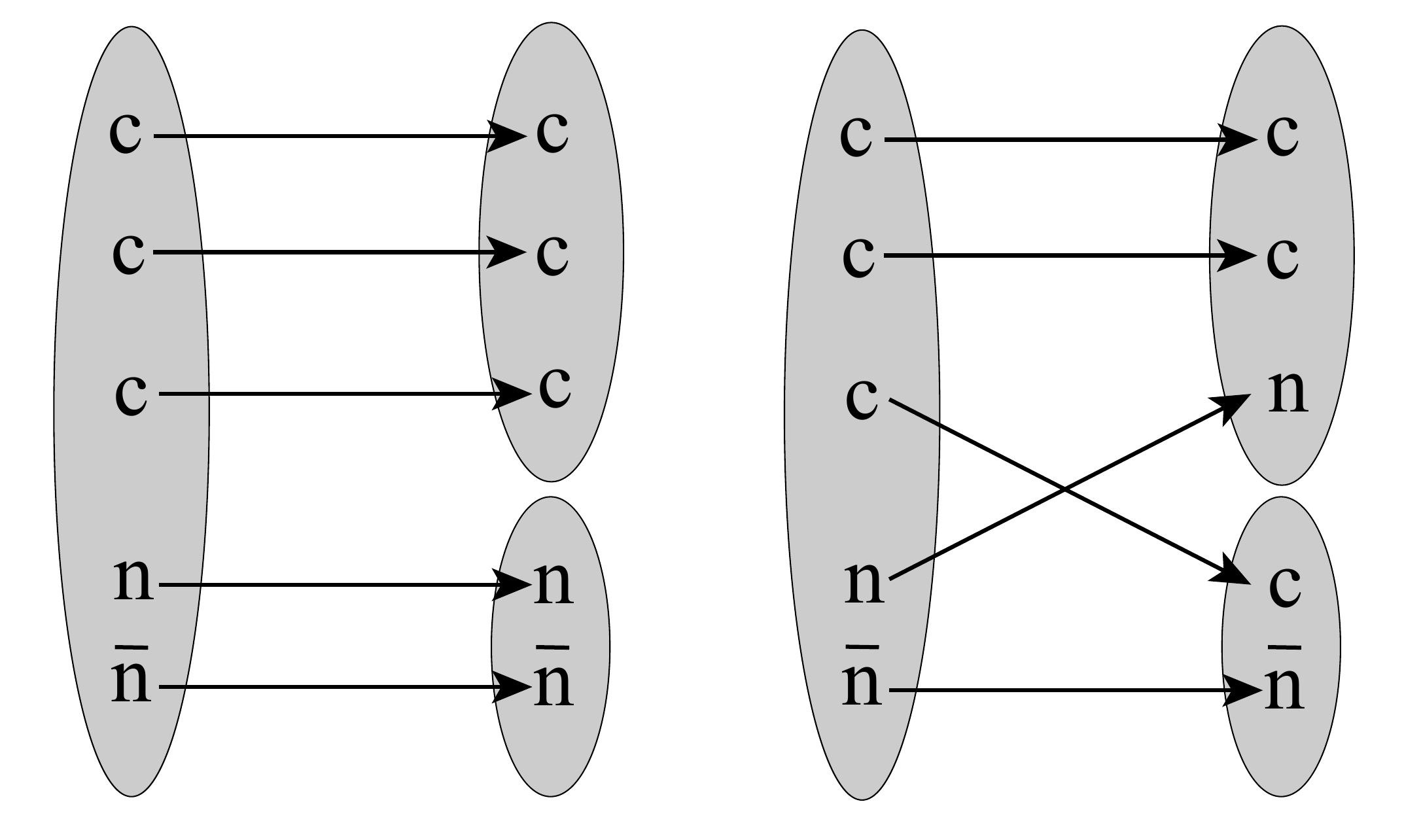}
\caption{The triply-heavy pentaquark state $cccn\bar{n}$ system has two decay combinations: the $1_{c}$ state of $cccn\bar{n}$ decays via $ccc \otimes n\bar{n}$, or $cccn\bar{n}$ transforms through the exchange of quarks into the form $ccnc\bar{n}$, which then decays through $ccn \otimes c\bar{n}$ in its $1_{c}$ state.}
\label{fig:fig2}
\end{figure}

After the scattering states are excluded, for the remaining compact $cccn\bar{n}$ pentaquark states, there exist two decay combinations of $ccc \otimes n\bar{n}$ and $ccn \otimes c\bar{n}$, as shown in the Fig.\ref{fig:fig2}.

When the $J^{P}$ quantum number is $1/2^{-}$, there exist three states with masses of 5.741 GeV, 5.827 GeV, and 5.963 GeV. In the case of the $ccc \otimes n\bar{n}$ type decay, the resulting baryon in the $1_{c}$ coupling can only be $\Omega_{ccc}$ of $J=3/2$, and the corresponding meson can be $\rho$ or $\omega$ of $J=1$. For the two possible decay channels, their decay coefficients satisfy the following relation:

\begin{equation}
\gamma_{\Omega_{ccc}\omega}=\gamma_{\Omega_{ccc}\rho}=\gamma_{\Omega_{ccc}\pi}.
\label{eq:eq15}
\end{equation}

Using the previous Eq.(\ref{eq:eq12}) and Eq.(\ref{eq:eq13}), we can get the partial decay width of each state on a specific decay channel. By choosing the channel with the largest partial decay width as a reference, we can obtain the partial decay width ratios for the states at 5.741 GeV, 5.827 GeV, and 5.963 GeV:

\begin{equation}
    \begin{aligned}
\frac{\Gamma(cccn\bar{n}\longrightarrow \Omega_{ccc}+\omega)}{\Gamma(cccn\bar{n}\longrightarrow \Omega_{ccc}+\rho)} = 0.970, \mathrm{\ at\ }5.741\mathrm{\ GeV}; \\
\frac{\Gamma(cccn\bar{n}\longrightarrow \Omega_{ccc}+\omega)}{\Gamma(cccn\bar{n}\longrightarrow \Omega_{ccc}+\rho)} = 0.983, \mathrm{\ at\ }5.827\mathrm{\ GeV}; \\
\frac{\Gamma(cccn\bar{n}\longrightarrow \Omega_{ccc}+\omega)}{\Gamma(cccn\bar{n}\longrightarrow \Omega_{ccc}+\rho)} = 0.990, \mathrm{\ at\ }5.963\mathrm{\ GeV}.
    \end{aligned}
\end{equation}

For the $ccn\otimes c\bar{n}$ type decay, the resulting baryon in the decay includes $\Xi_{cc}(J=1/2)$ and $\Xi_{cc}^{\ast}(J=3/2)$. In addition, the resulting meson also includes two possible particle types, namely $D$ and $D^{\ast}$.

When $J^{P}=1/2^{-}$, according to the color-spin wave functions in the appendix's $QQnQ\bar{n}$ section, the $ccc\otimes n\bar{n}$ state coupled with $8_{c}$ transforms via quark exchange to the $ccnc\bar{n}$ state coupled with $1_{c}$. If decay occurs in the form of $ccn\otimes c\bar{n}$, there exist three decay channels: $\Xi_{cc}\otimes D$, $\Xi_{cc}\otimes D^{\ast}$, and $\Xi_{cc}^{\ast}\otimes D^{\ast}$. The momentum $k$ can be calculated using Eq.(\ref{eq:eq13}) based on the masses of $\Xi_{cc}$, $\Xi_{cc}^{\ast}$, $D$, $D^{\ast}$, and the $cccn\bar{n}$ obtained from the MIT bag model.

According to the previously defined manner of $\gamma_{i}$, we can obtain the following relation:

\begin{equation}
\gamma_{\Xi_{cc} D}=\gamma_{\Xi_{cc} D^{\ast}}=\gamma_{\Xi_{cc}^{\ast} D}=\gamma_{\Xi_{cc}^{\ast} D^{\ast}}.
\end{equation}

The obtained decay coefficients $\gamma_{i}$ for each decay channel, along with the momentum $k$ coefficients, are inserted into Eq.(\ref{eq:eq12}). Through calculation, the partial decay widths for each decay channel can be obtained. For each compact pentaquark state, the decay channel with the largest decay width is chosen as the standard. By calculating the partial decay width ratios of other decay channels to the maximum partial decay width, we obtain the partial decay width ratios for each decay channel.

For $J^{P}=1/2^{-}$, using the decay width of $\Xi_{cc}D$ as the standard, the decay width ratios for the pentaquark state with mass of 5.741 GeV are:

\begin{equation}
    \begin{aligned}
\frac{\Gamma(cccn\bar{n}\longrightarrow \Xi_{cc}^{\ast}+D^{\ast})}{\Gamma(cccn\bar{n}\longrightarrow \Xi_{cc}+D)} =0.010,\\
\frac{\Gamma(cccn\bar{n}\longrightarrow \Xi_{cc}+D^{\ast})}{\Gamma(cccn\bar{n}\longrightarrow \Xi_{cc}+D)} =0.063.
    \end{aligned}
\end{equation}

 For the pentaquark state with 5.827 GeV, using the decay width of $\Xi_{cc}D^{\ast}$ as the standard, the partial decay width ratios are:

\begin{equation}
    \begin{aligned}
\frac{\Gamma(cccn\bar{n}\longrightarrow \Xi_{cc}^{\ast}+D^{\ast})}{\Gamma(cccn\bar{n}\longrightarrow \Xi_{cc}+D^{\ast})} =0.008,\\
\frac{\Gamma(cccn\bar{n}\longrightarrow \Xi_{cc}+D)}{\Gamma(cccn\bar{n}\longrightarrow \Xi_{cc}+D^{\ast})} =0.055.
    \end{aligned}
\end{equation}

 Finally, for the state at 5.963 GeV, using the decay width of $\Xi_{cc}^{\ast}D^{\ast}$ as the standard, the partial decay width ratios for each decay channel are:

\begin{equation}
    \begin{aligned}
\frac{\Gamma(cccn\bar{n}\longrightarrow \Xi_{cc}+D^{\ast})}{\Gamma(cccn\bar{n}\longrightarrow \Xi_{cc}^{\ast}+D^{\ast})} =0.005,\\
\frac{\Gamma(cccn\bar{n}\longrightarrow \Xi_{cc}+D)}{\Gamma(cccn\bar{n}\longrightarrow \Xi_{cc}^{\ast}+D^{\ast})} =0.033.
    \end{aligned}
\end{equation}

For $J^{P}=3/2^{-}$, the decay channels for the decay $ccn\otimes c\bar{n}$ include three possible combinations: $\Xi_{cc}^{\ast}\otimes D$, $\Xi_{cc}\otimes D^{\ast}$, and $\Xi_{cc}^{\ast}\otimes D^{\ast}$. Following the same calculation process, using $\Xi_{cc}^{\ast}\otimes D^{\ast}$ as the comparative standard, we can obtain the following results:

\begin{equation}
    \begin{aligned}
\frac{\Gamma(cccn\bar{n}\longrightarrow \Xi_{cc}+D^{\ast})}{\Gamma(cccn\bar{n}\longrightarrow \Xi_{cc}^{\ast}+D^{\ast})} =0.204,\\
\frac{\Gamma(cccn\bar{n}\longrightarrow \Xi_{cc}^{\ast}+D)}{\Gamma(cccn\bar{n}\longrightarrow \Xi_{cc}^{\ast}+D^{\ast})} =0.708.
    \end{aligned}
\end{equation}

Table \ref{table:tab2} shows the partial decay width ratios of all decay channels for compact pentaquark states with $cccn\bar{n}$ structure.

\renewcommand{\tabcolsep}{0.3cm}
\renewcommand{\arraystretch}{1.1}
\begin{table*}[!htbp]
    \caption{The partial decay width ratios for the decays of the pentaquark configurations $ccc\otimes n\bar{n}$ and $ccn\otimes c\bar{n}$.The unit of mass is GeV.}
    \begin{tabular}{cc|ccc|cccc}
        \hline\hline
         $ $&$ $ &$ccc \otimes n\bar{n}$ &$ $ &$ $ &$ccn\otimes c\bar{n}$ &$ $ &$ $ &$ $\\
         $J^{P}$ &Mass &$\Omega_{ccc}\rho $ &$\Omega_{ccc}\omega $ &$\Omega_{ccc}\pi$ &$\Xi_{cc} ^{*}D^{*}$ &$\Xi_{cc}^{*} D$
         &$\Xi_{cc}D ^{*}$ &$\Xi_{cc}D$
         \\ \hline
        ${{1/2}^{-}}$
            &5.741  &1   &0.970  &$ $      &0.010   &$ $    &0.063  &1       \\
            &5.827  &1   &0.983  &$ $      &0.008   &$ $   &1       &0.055   \\
            &5.963  &1   &0.990  &$ $      &1       &$ $   &0.005   &0.033   \\
        ${{3/2}^{-}}$
            &5.854  &0   &0      &1        &1       &0.708   &0.204 &$ $     \\
        \hline\hline
    \end{tabular}
    \label{table:tab2}
\end{table*}

It can be seen from the results in Table \ref{table:tab1} that for the $cccn\bar{n}$ system with $J^{P}=1/2^{-}$, none of the three states is a scattering state. When the $cccn\bar{n}$ system undergoes decay in the $ccc\otimes n\bar{n}$ manner, the $J^{P}=1/2^{-}$ decay channels include $\Omega_{ccc}\omega$ and $\Omega_{ccc}\rho$, while the $J^{P}=3/2^{-}$ decay channel is $\Omega_{ccc}\pi$. There is no shared decay channel between these two cases.

When decayed in the $ccn\otimes c\bar{n}$ configuration, the three states with $J^{P}$ quantum number $1/2^{-}$ have the same decay channels, but the dominant decay channels are completely different. The dominant decay channels are those in Table \ref{table:tab2} with a partial decay width ratio equal to 1.

The decay channel $\Omega_{ccc}\pi$ in the $ccc\otimes n\bar{n}$ configuration is forbidden when $J^{P}=1/2^{-}$, as it would violate the conservation of angular momentum. If the triply-heavy pentaquark state $cccn\bar{n}$ were to decay into the $\Omega_{ccc}\pi$ channel in the $ccc\otimes n\bar{n}$ configuration, the resulting decay products would generate an orbital angular momentum of $L=1$ between each other. While this would satisfy the conservation of angular momentum, the presence of orbital angular momentum would violate parity conservation after decay. Therefore, the $\Omega_{ccc}\pi$ decay channel is disallowed. The same holds true for other forbidden decay channels.

\subsection{$ccbn\bar{n}$ system}

Next, we investigate the decay channels present in the structure of the compact pentaquark states with flavor composition $ccbn\bar{n}$. Using the specific color-spin wave functions of the $QQQn\bar{n}$ type and the obtained eigenvectors, we differentiate between the scattering states and the compact pentaquark states for $ccbn\bar{n}$, as shown in Table \ref{table:tab3}.

\renewcommand{\tabcolsep}{0.4cm}
\renewcommand{\arraystretch}{1.1}
\begin{table*}[!htbp]
    \caption{Masses, bag radius, eigenvectors, and scattering states of the triply-heavy pentaquark state $ccbn\bar{n}$ system at each $J^{P}$ quantum number. The unit of mass is GeV.}
    \begin{tabular}{c|cccc}
        \hline\hline
          &$R_{0}=5.338\,\mathrm{GeV^{-1}}$ &  &  &  \\
        State &$J^{P}$ &Mass &Eigenvector &Scattering state \\ \hline
        $ccbn\bar{n}$
            &${1/2}^{-}$    &8.699   &(-0.081, -0.004, -0.068, -0.062, -0.001, -0.001, 0.001, 0.993)  &$\Omega_{ccb}\pi$\\
            &$ $            &9.087   &(-0.048, -0.437, -0.312, 0.131, -0.027, -0.575, 0.600, -0.019)   &$    $\\
            &$ $            &9.109   &(-0.079, -0.073, 0.035, 0.074, 0.254, -0.668, -0.686, 0.001)    &$ $\\
            &$ $    &9.144   &(0.153, 0.162, 0.859, -0.084, 0.108, -0.292, 0.323, 0.066)  &$    $\\
            &$ $            &9.172   &(-0.714, 0.088, -0.009, -0.676, 0.050, -0.068, 0.084, -0.100)   &$    $\\
            &$ $            &9.203   &(0.596, 0.134, 0.200, -0.621, -0.373, -0.237, -0.068, -0.004)    &$ $\\
            &$ $    &9.324   &(0.308, -0.309, -0.081, -0.317, 0.810, 0.206, 0.059, -0.001)  &$    $\\ &$ $            &9.299   &(-0.032, -0.810, 0.335, -0.149, -0.353, 0.187, -0.222, 0.008)    &$ $\\
            &${3/2}^{-}$    &8.712   &(0.078, -0.041, -0.002, -0.040, 0, -0.995, 0)                 &$\Omega_{ccb}^{\ast}\pi$\\
            &$ $            &9.113   &(-0.047, -0.015, 0.151, -0.019, 0.201, -0.003, 0.966) &$\Omega_{ccb}\rho/\omega $\\
            &$ $            &9.124   &(0.031, 0.009, -0.089, 0.012, -0.971, 0.002, 0.218) &$\Omega_{ccb}^{\ast}\rho/\omega$\\
            &$ $            &9.181   &(0.715, -0.679, -0.093, -0.093, 0.031, 0.088, 0.031)                 &$ $\\
            &$ $            &9.214   &(0.584, 0.653, -0.417, 0.209, 0.086, 0.011, 0.090) &$ $\\
            &$ $            &9.253   &(-0.165, -0.277, -0.191, 0.925, 0.027, -0.039, 0.030)  &$  $\\
            &$ $            &9.313   &(0.333, 0.184, 0.866, 0.299, -0.084, 0.005, -0.093)  &$  $\\
            &${5/2}^{-}$    &9.126            &(0, 1)              &$\Omega_{ccb}^{\ast}\rho/\omega$\\
            &$ $    &9.281            &(-1, 0)              &$ $\\
        \hline\hline
    \end{tabular}
    \label{table:tab3}
\end{table*}

\begin{figure}
\centering
\includegraphics[width=0.45\textwidth]{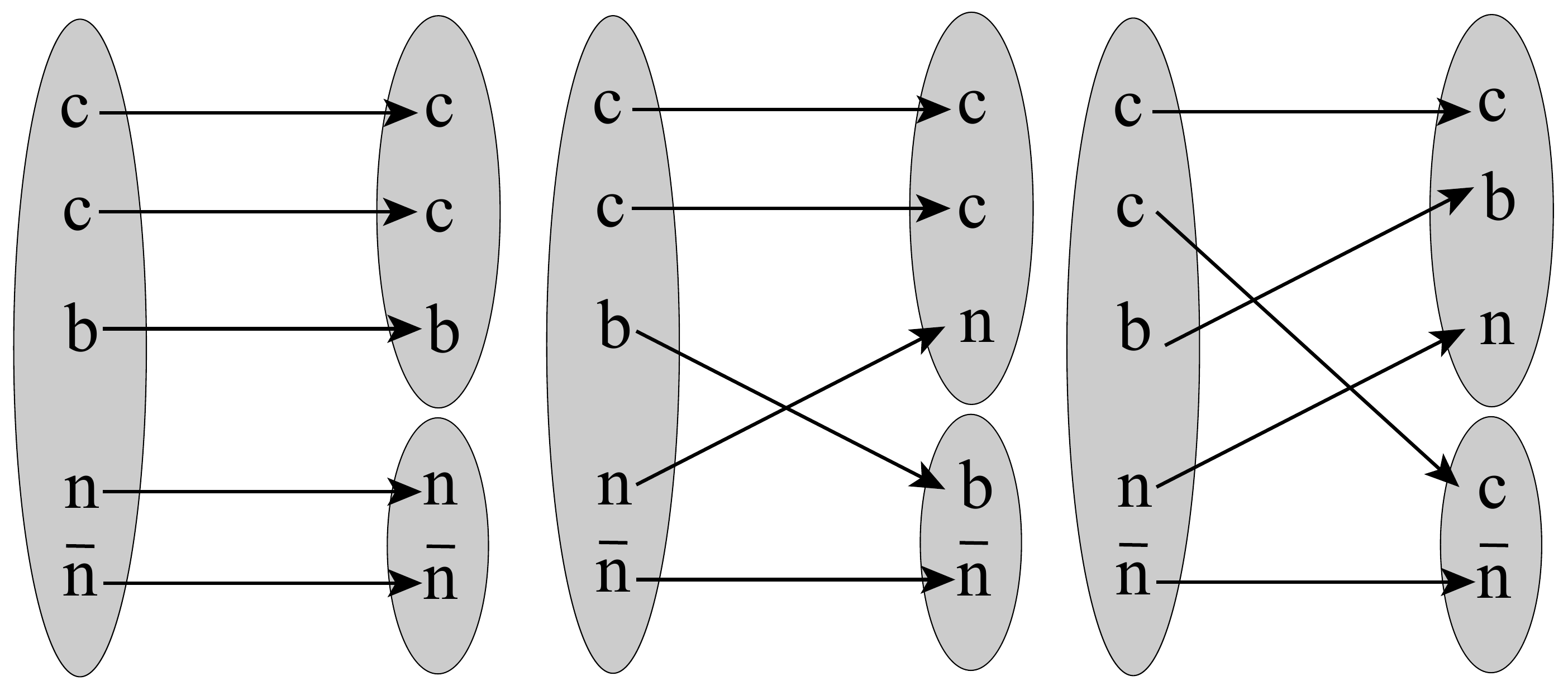}
\caption{For the triply-heavy pentaquark state $ccbn\bar{n}$, there are three decay combinations: the $1_{c}$ state of $ccbn\bar{n}$ decays via $ccb \otimes n\bar{n}$, or $ccbn\bar{n}$ transforms through the exchange of quarks into the $1_{c}$ states of $ccnb\bar{n}$ and $cbnc\bar{n}$, and subsequently decays accordingly.}
\label{fig:fig3}
\end{figure}

After scattering states are excluded, decay widths of the remaining compact pentaquark states are analyzed. The approach for handling the $ccbn\bar{n}$ system is similar to that of the $cccn\bar{n}$ system. The difference is that the decay channels of the compact pentaquark state $ccbn\bar{n}$ involves three configurations of  $ccb\otimes n\bar{n}$, $ccn\otimes b\bar{n}$, and $cnb\otimes c\bar{n}$, as shown in Fig.\ref{fig:fig3}.

The color-spin wave function of the $ccb\otimes n\bar{n}$ type has three $1_{c}$ states each for $J^{P}=1/2^{-}$ and $J^{P}=3/2^{-}$. Considering that mesons of the $n\bar{n}$ configuration include particles $\omega$ and $\rho$ with spin 1, then there are six decay channels. For these compact triply-heavy pentaquark states, if they decay into the $1_{c}$ state of $ccb\otimes n\bar{n}$, the
decay coefficients of their six decay channels satisfy the following relation:

\begin{equation}
\gamma_{\Omega_{ccb}^{\ast}\omega} = \gamma_{\Omega_{ccb}^{\ast}\rho}=
\gamma_{\Omega_{ccb}^{\ast}\pi}=
\gamma_{\Omega_{ccb}\omega} = \gamma_{\Omega_{ccb}\rho}=
\gamma_{\Omega_{ccb}\pi}.
\end{equation}

In the same case, the decay coefficients of decay channels in the $ccn\otimes b\bar{n}$ and $cbn\otimes c\bar{n}$ configurations also satisfy a similar relation:

\begin{equation}
    \begin{aligned}
&\gamma_{\Xi_{cc}^{\ast}B^{\ast}}=\gamma_{\Xi_{cc}B ^{\ast}}=\gamma_{\Xi_{cc}^{\ast}B}=\gamma_{\Xi_{cc}B},\\
\gamma_{\Xi_{bc}^{\ast}D^{\ast}}=&\gamma_{\Xi_{bc}^{'}D ^{\ast}}=\gamma_{\Xi_{bc}D ^{\ast}}
=\gamma_{\Xi_{bc}^{'}D}=\gamma_{\Xi_{bc}^{\ast}D}=\gamma_{\Xi_{bc}D}.\\
    \end{aligned}
\end{equation}

Using the previous Eq.(\ref{eq:eq12}) and Eq.(\ref{eq:eq13}), we can obtain the partial decay width ratios of the three configurations, as shown in Table \ref{table:tab4} and Table \ref{table:tab5}.

\renewcommand{\tabcolsep}{0.35cm}
\renewcommand{\arraystretch}{1.1}
\begin{table*}[!htbp]
    \caption{The partial width ratios for the decays of the pentaquark configurations $ccb\otimes n\bar{n}$ and $ccn\otimes b\bar{n}$. The unit of mass is GeV.}
    \begin{tabular}{cc|cccccc|cccc}
        \hline\hline
         $ $&$ $&$ccb\otimes n\bar{n}$&$ $&$ $&$ $&$ $&$ $&$ccn\otimes b\bar{n}$&$ $&$ $&$ $ \\
         $J^{P}$&Mass&$\Omega_{ccb}^{\ast}\rho$&$\Omega_{ccb}^{\ast}\omega$&$\Omega_{ccb}^{\ast}\pi$&$\Omega_{ccb}\rho$&$\Omega_{ccb}\omega$&$\Omega_{ccb}\pi$&$\Xi_{cc}^{\ast}B^{\ast}$&$\Xi_{cc}^{\ast}B$&$\Xi_{cc}B ^{\ast}$&$\Xi_{cc}B$\\ \hline
        ${{1/2}^{-}}$
        &9.087 &0.865 &0.847   &$ $    &1 &0.982&0.002 &0.112 &$ $ &0.029     &1 \\
        &9.109 &0.897 &0.881   &$ $    &1 &0.984&0 &0.003 &$ $ &0.204    &1 \\
        &9.144 &0.776 &0.764   &$ $    &1 &0.986&0.063 &0.027 &$ $ &0.541 &1 \\
        &9.172 &0.304 &0.300   &$ $    &0.484 &0.478&1 &0.361 &$ $ &1  &0.074 \\
        &9.203 &1 &0.988   &$ $    &0.085 &0.084&0.004 &0.001 &$ $ &1   &0.731 \\
        &9.299 &0.685 &0.680   &$ $    &1 &0.992&0.002 &1 &$ $ &0.821   &0.005 \\
        &9.324 &1 &0.988   &$ $    &0.084 &0.083&0 &1 &$ $ &0.001   &0.054 \\
        ${{3/2}^{-}}$
        &9.181 &0.083 &0.082 &1  &0.085 &0.084 &$ $ &0.246 &0.760 &1   &$ $  \\
        &9.214 &0.870 &0.860 &0.021  &1 &0.990 &$ $ &0.002 &0.012  &1  &$ $  \\
        &9.253 &0.359 &0.355 &1  &0.439 &0.435 &$ $ &0.381  &1 &0.065 &$ $  \\
        &9.313 &0.782 &0.776 &0.003  &1 &0.993 &$ $ &1  &0.218 &0.087 &$ $  \\
        ${{5/2}^{-}}$
        &9.281 &$ $ &$ $ &$ $  &$ $ &$ $ &$ $ &1 &$ $   &$ $ &$ $  \\
        \hline\hline
    \end{tabular}
    \label{table:tab4}
\end{table*}

\renewcommand{\tabcolsep}{0.2cm}
\renewcommand{\arraystretch}{1.1}
\begin{table*}[!htbp]
    \caption{The partial width ratios for the decays of the pentaquark configurations $cbn\otimes c\bar{n}$. The unit of mass is GeV.}
    \begin{tabular}{cc|cccccc}
        \hline\hline
         $ $&$ $&$cbn\otimes c\bar{n}$ &$ $ &$ $ &$ $ &$ $ &$ $ \\
         $J^{P}$&Mass&$\Xi_{bc}^{\ast}D^{\ast}$&$\Xi_{bc}^{\ast}D$&$\Xi_{bc}^{'}D^{\ast}$&$\Xi_{bc}D^{\ast}$&$\Xi_{bc}^{'}D$  &$\Xi_{bc}D$
         \\ \hline
        ${{1/2}^{-}}$
        &9.087 &0.005 &$ $ &0.010 &0.015 &1       &0.391 \\
        &9.109 &0.017 &$ $ &0.122 &0.021 &0.899   &1 \\
        &9.144 &0.033 &$ $ &0.026 &0.262 &1       &0.019 \\
        &9.172 &0.137 &$ $ &0.201 &0.206 &0.033   &1 \\
        &9.203 &0.010 &$ $ &1     &0.031 &0.048   &0.002 \\
        &9.299 &0.919 &$ $ &0.164 &1     &0.012   &0.070 \\
        &9.324 &1     &$ $ &0.049 &0.171 &0.054   &0.049 \\

        ${{3/2}^{-}}$
        &9.181 &0.292 &1     &0.007 &0.245 &$ $ &$ $  \\
        &9.214 &0.147 &0.035 &1     &0.618 &$ $ &$ $  \\
        &9.253 &0.939 &0.046 &0.182 &1     &$ $ &$ $  \\
        &9.313 &1     &0.140 &0.111 &0.374 &$ $ &$ $  \\
        ${{5/2}^{-}}$
        &9.281 &1 &$ $ &$ $ &$ $  &$ $ &$ $   \\
        \hline\hline
    \end{tabular}
    \label{table:tab5}
\end{table*}

In the states of $J^{P}=1/2^{-}$, the three states with masses of 9.087 GeV, 9.109 GeV and 9.144 GeV are primarily dominated by the decay channel $\Omega_{ccb}\rho$. However, as the mass increases, the dominant decay channel begins to change. The dominant decay of the 9.172 GeV state is the decay channel $\Omega_{ccb}\pi$, because the eigenvector of the decay channel $\Omega_{ccb}\pi$ in 9.172 GeV is larger than the eigenvector of the other decay channels. In addition, the other states are more inclined to $\rho/\omega$ meson decays.

Combined with the result of $J^{P}=3/2^{-}$, we find that other decay channels are also possible when the $\pi$ meson decay channel is dominant. However, when the decay channel of the meson product is $\rho/\omega$, almost no decay of the $\pi$ meson occurs. This indicates that decay is more likely to occur through channels with higher product mass.

For the two configurations of $ccn\otimes b\bar{n}$ and $cbn\otimes c\bar{n}$, we can obviously see that as the mass increases, the product mass of the dominant decay channel also increases slowly, which is reflected in the angular momentum of the decay product. Moreover, the partial decay width ratios of these dominant decay channels are in most cases significantly larger than those of other possible decay channels.

\subsection{$bbcn\bar{n}$ system}

For the $bbcn\bar{n}$ system, the coefficients corresponding to the $1_{c}$ state in the eigenvectors calculated can differentiate scattering states among states for various $J^{P}$ quantum numbers, as presented in Table \ref{table:tab6}.

\renewcommand{\tabcolsep}{0.4cm}
\renewcommand{\arraystretch}{1.1}
\begin{table*}[!htbp]
    \caption{Masses, bag radius, eigenvectors, and scattering states of the triply-heavy pentaquark state $bbcn\bar{n}$ system at each $J^{P}$ quantum number. The unit of mass is GeV.}
    \begin{tabular}{c|cccc}
        \hline\hline
        &$R_{0}=5.164\,\mathrm{GeV^{-1}}$ &  &  &  \\
        State &$J^{P}$ &Mass &Eigenvector &Scattering state \\ \hline
        $bbcn\bar{n}$
            &${1/2}^{-}$    &12.034   &(-0.027, -0.003, -0.097, -0.046, -0.007, 0.001, 0.001, 0.994)  &$\Omega_{bbc}\pi$\\
            &$ $            &12.443   &(0.007, -0.009, -0.254, -0.086, -0.290, 0.602, 0.693, -0.031)   &$    $\\
            &$ $            &12.453   &(-0.032, -0.109, 0.075, 0.039, 0.109, -0.703, 0.688, 0.009)    &$ $\\
            &$ $            &12.479   &(-0.079, -0.101, -0.882, -0.247, -0.150, -0.284, -0.175, -0.101) &$    $\\
            &$ $            &12.574   &(0.904, 0.256, -0.140, 0.273, -0.122, -0.092, 0.007, 0.023)   &$    $\\
            &$ $            &12.584   &(-0.210, -0.253, -0.256, 0.894, 0.137, 0.089, -0.003, 0.011)   &$    $\\
            &$ $            &12.647   &(-0.053, -0.299, 0.234, 0.122, -0.890, -0.176, -0.121, 0.020)   &$    $\\
            &$ $            &12.685   &(0.358, -0.871, 0.048, -0.197, 0.237, 0.122, -0.029, 0.004)   &$    $\\
            &${3/2}^{-}$    &12.047   &(0.23, 0.043, 0.004, -0.062, 0, -0.997, 0)                 &$\Omega_{bbc}^{\ast}\pi$\\
            &$ $    &12.446            &(-0.018, 0.067, 0.204, -0.066, -0.217, 0.008, 0.950)              &$\Omega_{bbc}\rho/\omega$\\
            &$ $    &12.462            &(-0.008, 0.034, 0.076, -0.032, -0.966, 0.004, -0.242)           &$\Omega_{bbc}^{\ast}\rho/\omega$\\
            &$ $            &12.503   &(-0.088, 0.777, 0.199, -0.568, 0.093, 0.067, -0.118)
            &$  $\\
            &$ $            &12.596   &(0.091, -0.611, 0.251, -0.743, 0.035, 0.023, -0.053)   &$ $\\
            &$ $    &12.619            &(0.878, 0.120, -0.430, -0.143, -0.051, 0.033, 0.078)           &$ $\\
            &$ $    &12.645            &(0.461, 0.041, 0.816, 0.311, 0.083, -0.004, -0.128)            &$ $\\
            &${5/2}^{-}$    &12.463            &(0, 1)              &$\Omega_{bbc}^{\ast}\rho/\omega$\\
            &$  $    &12.619            &(1, 0)             &$ $\\
        \hline\hline
    \end{tabular}
    \label{table:tab6}
\end{table*}

After excluding the scattering states, for the remaining compact pentaquark states in $bbcn\bar{n}$, there exist three decay configurations: $bbc\otimes n\bar{n}$, $bbn\otimes c\bar{n}$, and $bcn\otimes b\bar{n}$. The decay coefficients for the different decay channels within these three decay configurations satisfy the following relations:

\begin{equation}
    \begin{aligned}
\gamma_{\Omega_{bbc}^{\ast}\omega}=&\gamma_{\Omega_{bbc}^{\ast}\rho}=\gamma_{\Omega_{bbc}^{\ast}\pi}=\gamma_{\Omega_{bbc}\omega}=\gamma_{\Omega_{bbc}\rho}=\gamma_{\Omega_{bbc}\pi},\\       &\gamma_{\Xi_{bb}^{\ast}D^{\ast}}=\gamma_{\Xi_{bb}D ^{\ast}}=\gamma_{\Xi_{bb}^{\ast}D}=\gamma_{\Xi_{bb}D} ,\\       \gamma_{\Xi_{bc}^{\ast}B^{\ast}}=&\gamma_{\Xi_{bc}^{'}B ^{\ast}}=\gamma_{\Xi_{bc}B^{\ast}}
=\gamma_{\Xi_{bc}^{\ast}B}=\gamma_{\Xi_{bc}^{'}B}=\gamma_{\Xi_{bc}B}.
    \end{aligned}
\end{equation}

For each compact pentaquark of a $bbcn\bar{n}$ system, the partial decay width of each decay channel can be obtained by Eq.(\ref{eq:eq12}). The partial decay channel with the largest decay width is selected for each state in the specific configuration as a reference, thus obtaining the partial decay width ratios for the corresponding decay channel of each state. We list them in Tables \ref{table:tab7} and \ref{table:tab8}.

\renewcommand{\tabcolsep}{0.35cm}
\renewcommand{\arraystretch}{1.1}
\begin{table*}[!htbp]
    \caption{The partial width ratios for the decays of the pentaquark configurations $bbc\otimes n\bar{n}$ and $bbn\otimes c\bar{n}$. The unit of mass is GeV.}
    \begin{tabular}{cc|cccccc|cccc}
        \hline\hline
         $ $&$ $&$bbc\otimes n\bar{n}$&$ $&$ $&$ $&$ $&$ $&$bbn\otimes c\bar{n}$&$ $&$ $&$ $  \\
         $J^{P}$&Mass&$\Omega_{bbc}^{\ast}\rho$&$\Omega_{bbc}^{\ast}\omega$&$\Omega_{bbc}^{\ast}\pi$&$\Omega_{bbc}\rho$&$\Omega_{bbc}\omega$&$\Omega_{bbc}\pi$&$\Xi_{bb}^{\ast}D^{\ast}$&$\Xi_{bb}^{\ast}D$&$\Xi_{bb}D ^{\ast}$&$\Xi_{bb}D$\\ \hline
        ${{1/2}^{-}}$
            &12.443 &0.713 &0.704 &$ $ &1  &0.988 &0.003    &0.010  &$ $  &0.034  &1 \\
            &12.453 &0.985 &0.973 &$ $ &1  &0.989 &0.0002   &0.047  &$ $  &0.909  &1 \\
            &12.479 &1 &0.989 &$ $ &0.398  &0.394 &0.185   &0.071  &$ $&0.043    &1 \\
            &12.574 &1 &0.992 &$ $ &0.006  &0.006 &0.086   &0.615 &$ $  &1   &0.120\\
            &12.584 &1 &0.992 &$ $ &0.002  &0.002 &0.020   &0.051 &$ $  &1   &0.012\\
            &12.647 &1 &0.993 &$ $ &0.493  &0.490 &0.016   &1 &$ $  &0.090   &0.039\\
            &12.685 &1 &0.994 &$ $ &0.058  &0.057 &0.001   &1 &$ $ &0.113    &0.075\\
        ${{3/2}^{-}}$
            &12.503 &0.583 &0.578 &0.430  &1   &0.991 &$ $  &0.022  &1  &0.085    &$ $  \\
            &12.596 &0.425 &0.422 &0.233  &1   &0.993 &$ $  &0.300  &1  &0.030    &$ $  \\
            &12.619 &0.416 &0.413 &0.217  &1   &0.993 &$ $  &1  &0.656   &0.009   &$ $  \\
            &12.645 &0.409 &0.407 &0.012  &1   &0.994 &$ $  &1  &0.204  &0.085    &$ $  \\
        ${{5/2}^{-}}$
            &12.619 &$ $ &$ $ &$ $  &$ $   &$ $ &$ $  &1  &$ $ &$ $     &$ $  \\
        \hline\hline
    \end{tabular}
    \label{table:tab7}
\end{table*}

\renewcommand{\tabcolsep}{0.3cm}
\renewcommand{\arraystretch}{1.1}
\begin{table*}[!htbp]
    \caption{The partial width ratios for the decays of the pentaquark configurations $bcn\otimes b\bar{n}$. The unit of mass is GeV.}
    \begin{tabular}{cc|cccccc}
        \hline\hline
         $ $&$ $&$bcn\otimes b\bar{n}$ &$ $ &$ $ &$ $ &$ $ &$ $ \\
         $J^{P}$&Mass&$\Xi_{bc}^{\ast}B^{\ast}$&$\Xi_{bc}^{\ast}B$&$\Xi_{bc}^{'}B^{\ast}$&$\Xi_{bc}B^{\ast}$&$\Xi_{bc}^{'}B$&$\Xi_{bc}B$
         \\ \hline
        ${{1/2}^{-}}$
         &12.443 &0.066 &$ $ &0.139  &0.018 &0.103 &1 \\
         &12.453 &0.028 &$ $ &0.043  &0.170 &1 &0.538 \\
         &12.479 &0.087 &$ $ &0.342  &0.254 &1 &0.058 \\
         &12.574 &0.013 &$ $ &1   &0.287 &0.115&0.530 \\
         &12.584 &0.006 &$ $ &1  &0.112 &0.870 &0.821 \\
         &12.647 &0     &$ $ &0.022   &1  &0.004&0.330 \\
         &12.685 &1     &$ $ &0.002  &0.084 &0.084  &0.075 \\
        ${{3/2}^{-}}$
         &12.503 &0.372 &0.671 &0.628 &1 &$ $  &$ $ \\
         &12.596 &0.518 &0.783 &0.435 &1 &$ $  &$ $ \\
         &12.619 &0.618 &1     &0.628 &0.575 &$ $  &$ $ \\
         &12.645 &1     &0.021 &0.045 &0.422 &$ $  &$ $ \\
        ${{5/2}^{-}}$
         &12.619 &1   &$ $ &$ $ &$ $ &$ $ &$ $  \\
        \hline\hline
    \end{tabular}
    \label{table:tab8}
\end{table*}

The structures of the $bbcn\bar{n}$ system and the $ccbn\bar{n}$ system are very similar. Since the masses of the $bbcn\bar{n}$ system are larger, they are more likely to decay. Therefore, as the mass of the $bbcn\bar{n}$ system increases, the characteristics of the decay channels become more obvious.

When the $bbcn\bar{n}$ system decays through the $bbc\otimes n\bar{n}$ configuration, the two states of 12.443 GeV and 12.453 GeV at $J^{P}=1/2^{-}$, the partial decay width ratios of the four decay channels $\Omega_{bbc}^{\ast}\rho$, $\Omega_{bbc}^{\ast}\omega$, $\Omega_{bbc}\rho$ and $\Omega_{bbc}\omega$ are all larger. However, the $bbcn\bar{n}$ system begins to tend to the dominant decay channels $\Omega_{bbc}^{\ast}\rho$ and  $\Omega_{bbc}^{\ast}\omega$ as the mass increases.
When $J^{P}=3/2^{-}$, the two decay channels $\Omega_{bbc}\rho$ and $\Omega_{bbc}\omega$ always dominate.

When the $bbcn\bar{n}$ system decays in two configurations of $bbn\otimes c\bar{n}$ and $bcn\otimes b\bar{n}$, it can be clearly found that the decay channel tends to increase the decay product mass with the increase of mass. Tables \ref{table:tab7} and \ref{table:tab8} show that the angular momentum of the decay products is increasing.

\subsection{$bbbn\bar{n}$ system}

Finally, turning our attention to the $bbbn\bar{n}$ system, similar to the earlier $cccn\bar{n}$ system, after excluding the scattering states using the obtained eigenvectors, we obtain Table \ref{table:tab9}.

\renewcommand{\tabcolsep}{0.6cm}
\renewcommand{\arraystretch}{1.1}
\begin{table*}[!htbp]
    \caption{ Masses, bag radius, eigenvectors, and scattering states of the triply-heavy pentaquark state $bbbn\bar{n}$ system at each $J^{P}$ quantum number. The unit of mass is GeV and the unit of bag radius is $\mathrm{ GeV^{-1} }$. }

    \begin{tabular}{c|ccccc}
        \hline\hline
        State &$J^{P}$ &$R_0$ &Mass &Eigenvector &Scattering state \\ \hline
        $bbbn\bar{n}$
            &${1/2}^{-}$    &5.032  &16.032   &(0.357,  -0.928,  0.111)  &$    $\\
            &$ $            &4.940  &15.957   &(-0.933, -0.355, 0.064)   &$    $\\
            &$ $            &4.996  &15.792   &(0.018,  0.126, 0.992)    &$\Omega_{bbb}\rho/\omega$\\
            &${3/2}^{-}$    &5.011  &15.796   &(0, 1, 0)                 &$\Omega_{bbb}\rho/\omega$\\
            &$ $            &4.972  &15.967   &(0.999, 0, 0.034)         &$    $\\
            &$ $            &4.846  &15.377   &(-0.035, 0, 0.999)        &$ \Omega_{bbb}\pi   $\\
            &${5/2}^{-}$    &5.011  &15.796            &(1)              &$\Omega_{bbb}\rho/\omega$\\
        \hline\hline
    \end{tabular}
    \label{table:tab9}
\end{table*}

The compact pentaquark states $bbbn\bar{n}$ have two decay configurations: $bbb\otimes n\bar{n}$ and $bbn\otimes b\bar{n}$. The decay coefficients for different decay channels satisfy the following relations:

\begin{equation}
    \begin{aligned}
        &\gamma_{\Omega_{bbb}\omega}=\gamma_{\Omega_{bbb}\rho}=\gamma_{\Omega_{bbb}\pi},\\
        &\gamma_{\Xi_{bb}^{\ast}B^{\ast}}=\gamma_{\Xi_{bb}^{\ast}B}=\gamma_{\Xi_{bb}B^{\ast}}=\gamma_{\Xi_{bb}B}.
    \end{aligned}
\end{equation}

According to the same computational method, we can list the decay partial width ratio for each decay channel of $bbbn\bar{n}$ as shown in Table \ref{table:tab10}.

\renewcommand{\tabcolsep}{0.3cm}
\renewcommand{\arraystretch}{1.1}
\begin{table*}[!htbp]
    \caption{The partial width ratios for the decays of the pentaquark configurations $bbb\otimes n\bar{n}$ and $bbn\otimes b\bar{n}$. The unit of mass is GeV.}
    \begin{tabular}{cc|ccc|cccc}
        \hline\hline
         $ $&$ $&$bbb\otimes n\bar{n}$&$ $&$ $&$bbn\otimes b\bar{n}$&$ $&$ $&$ $  \\ $J^{P}$&Mass&$\Omega_{bbb}\rho$&$\Omega_{bbb}\omega$&$\Omega_{bbb}\pi$
         &$\Xi_{bb}^{\ast}B^{\ast}$&$\Xi_{bb}^{\ast}B$&$\Xi_{bb}B^{\ast}$&$\Xi_{bb}B$\\ \hline
        ${{1/2}^{-}}$
            &15.957  &1 &0.995 &$ $ &0.023 &$ $   &1      &0.236 \\
            &16.032  &1 &0.995 &$ $ &1     &$ $   &0      &0.124 \\
        ${{3/2}^{-}}$
            &15.967  &0 &0     &1   &1     &0.608 &0.198  &$ $   \\
        \hline\hline
    \end{tabular}
    \label{table:tab10}
\end{table*}

In the decay process of the $bbb\otimes n\bar{n}$ configuration, there are distinct decay channels under each $J^{P}$ quantum number. However, in the $bbn\otimes b\bar{n}$ configuration, the angular momentum of decay products increases with the increase of mass of decay channels dominated by different states.

\section{summary}
\label{sec:summary}

This study comprehensively investigates the properties of the triply-heavy pentaquark state $QQQn\bar{n}$ within the framework of the MIT bag model. We provide a detailed characterization of the mass ranges for different systems of $QQQ n\bar{n}$. In this process, we observe that for the $cccn\bar{n}$ system, the mass range is approximately 5.7-6.0 GeV, while the $ccbn\bar{n}$ system has a mass range of 9.1-9.3 GeV. The mass for the $bbcn\bar{n}$ system falls within the 12.4-12.7 GeV range, and the $bbbn\bar{n}$ system has a mass of approximately 16.0 GeV. Since the masses of the triply-heavy pentaquarks calculated in this work are all above their respective baryon-meson thresholds, all of these pentaquarks are unstable in the two-body strong decay.

In addition to mass, we also systematically study the partial decay width ratios of decay channels of different configurations in each $QQQn\bar{n}$ system. Since angular momentum is conserved during decay, this limits some types of decay channels. If there is an orbital angular momentum $L=1$ between the two particles produced by the decay, then conservation of angular momentum can be satisfied. However, the parity will have an additional -1 related to the orbital angular momentum, resulting in non-conservation of parity for the entire decay process. Some decay channels are therefore prohibited.

After the scattering states are excluded, we find that as the mass of each configuration of the triply-heavy pentaquark state increases, the dominant decay channels (i.e., the decay channels with the partial decay width ratio equal to 1 in each state) tend to the states with the increase in angular momentum of the decay products. The partial decay width ratio of the final dominant decay channels will be much larger than other decay channels. This suggests that once the momentum required for decay is satisfied, the residual mass will be as much as possible in the form of decay products.

We hope that these conclusions can be verified in future experiments on the triply-heavy pentaquark state $QQQ n\bar{n}$.

\section*{ACKNOWLEDGEMENTS}
D. J. is supported by the National Natural Science Foundation of China under Grant No. 12165017.

\section*{Appendix A: Color and Spin Wave functions}
\label{apd:basis}
\setcounter{equation}{0}
\renewcommand{\theequation}{A\arabic{equation}}

\begin{equation}
    \begin{aligned}
        \phi_{1}^{P}&=|[(12)^{6}3]^{8}(4\bar{5})^{8} \rangle\\
        &=\frac{1}{4\sqrt{3}}\Big[2\bigl(rrgb\bar{r}+ggbr\bar{g}+bbrg\bar{b}-rrbg\bar{r}-bbgr\bar{b}\\
        &-ggrb\bar{g}\bigr)+\bigl(rgbr\bar{r}+grbr\bar{r}+gbrg\bar{g}+bgrg\bar{g}+brgb\bar{b}\\
        &+rbgb\bar{b}-brgr\bar{r}-rbgr\bar{r}-rgbg\bar{g}-grbg\bar{g}-gbrb\bar{b}\\
        &-bgrb\bar{b}+rggb\bar{g}+grgb\bar{g}+gbbr\bar{b}+bgbr\bar{b}+brrg\bar{r}\\
        &+rbrg\bar{r}-rgrb\bar{r}-grrb\bar{r}-gbgr\bar{g}-bggr\bar{g}-brbg\bar{b}\\
        &-rbbg\bar{b}\bigr)\Big],
    \end{aligned}
\end{equation}

\begin{equation}
    \begin{aligned}
        \phi_{2}^{P}&=|[(12)^{\bar{3}}3]^{8}(4\bar{5})^{8}\rangle\\
        &=\frac{1}{12}\Big[3\bigl(rgrb\bar{r}-grrb\bar{r}+rggb\bar{g}-grgb\bar{g}+gbgr\bar{g}\\
        &-bggr\bar{g}+gbbr\bar{b}-bgbr\bar{b}+brbg\bar{b}-rbbg\bar{b}\\
        &+brrg\bar{r}-rbrg\bar{r}\bigr)+2\bigl(rgbb\bar{b}-grbb\bar{b}+gbrr\bar{r}\\
        &-bgrr\bar{r}+brgg\bar{g}-rbgg\bar{g}\bigr)-\bigl(rgbr\bar{r}-grbr\bar{r}\\
        &+gbrg\bar{g}-bgrg\bar{g}+brgb\bar{b}-rbgb\bar{b}+rgbg\bar{g}\\
        &-grbg\bar{g}+gbrb\bar{b}-bgrb\bar{b}+brgr\bar{r}-rbgr\bar{r}\bigr)\Big],
    \end{aligned}
\end{equation}

\begin{equation}
    \begin{aligned}
        \phi_{3}^{P}&=|[(12)^{\bar{3}}3]^{1}(4\bar{5})^{1}\rangle\\
        &=\frac{1}{3\sqrt{2}}\Big[\bigl(grb-rgb+rbg-brg+bgr-gbr\bigr)r\bar{r}\\
        &+\bigl(grb-rgb+rbg-brg+bgr-gbr\bigr)g\bar{g}\\
        &+\bigl(grb-rgb+rbg-brg+bgr-gbr\bigr)b\bar{b}\Big].
    \end{aligned}
\end{equation}

\begin{equation}
    \begin{aligned}
    \chi_{1}^{P}&=|[(12)_{1}3]_{3/2}(4\bar{5})_{1}\rangle_{5/2},
    \chi_{2}^{P}=|[(12)_{1}3]_{3/2}(4\bar{5})_{1} \rangle_{3/2},\\
    \chi_{3}^{P}&=|[(12)_{1}3]_{3/2}(4\bar{5})_{0}\rangle_{3/2},
    \chi_{4}^{P}=|[(12)_{1}3]_{1/2}(4\bar{5})_{1} \rangle_{3/2},\\
    \chi_{5}^{P}&=|[(12)_{0}3]_{1/2} (4\bar{5})_{1}\rangle_{3/2},
    \chi_{6}^{P}=|[(12)_{1}3]_{3/2} (4\bar{5})_{1}\rangle_{1/2},\\
    \chi_{7}^{P}&=|[(12)_{1}3]_{1/2} (4\bar{5})_{1}\rangle_{1/2},
    \chi_{8}^{P}=|[(12)_{1}3]_{1/2}(4\bar{5})_{0}\rangle_{1/2},\\
    \chi_{9}^{P}&=|[(12)_{0}3]_{1/2}(4\bar{5})_{1}\rangle_{1/2},
    \chi_{10}^{P}=|[(12)_{0}3]_{1/2} (4\bar{5})_{0}\rangle_{1/2}.
    \end{aligned}
\end{equation}

\section*{Appendix B: The pentaquark Wave functions}
\label{apd:basis}
\setcounter{equation}{0}
\renewcommand{\theequation}{B\arabic{equation}}

The wave function of pentaquark $QQQn\bar{n}$.

a.$J^{P}=5/2^{-}:$
\begin{equation}
  \phi_{3}\chi_{1} = | [(12)^{\bar{3}}_{1}3]^{1}_{3/2} (4\bar{5})^{1}_{1} \rangle_{5/2}.
\end{equation}

b.$J^{P}=3/2^{-}:$
\begin{align}
  \frac{1}{\sqrt{2}}\big(\phi_{1}\chi_{5}-\phi_{2}\chi_{4}\big)=&\frac{1}{\sqrt{2}}\big(| [(12)^{6}_{0}3]^{8}_{1/2} (4\bar{5})^{8}_{1} \rangle_{3/2} \nonumber \\
  &-| [(12)^{\bar{3}}_{1}3]^{8}_{1/2} (4\bar{5})^{8}_{1} \rangle_{3/2}\big), \nonumber \\
  \phi_{3}\chi_{2} =& | [(12)^{\bar{3}}_{1}3]^{1}_{3/2} (4\bar{5})^{1}_{1} \rangle_{3/2}, \nonumber \\
  \phi_{3}\chi_{3} =& | [(12)^{\bar{3}}_{1}3]^{1}_{3/2} (4\bar{5})^{1}_{0} \rangle_{3/2}.
\end{align}

c.$J^{P}=1/2^{-}:$
\begin{align}
  \frac{1}{\sqrt{2}}\left(\phi_{1}\chi_{9}-\phi_{2}\chi_{7}\right)=&\frac{1}{\sqrt{2}}\big(| [(12)^{6}_{0}3]^{8}_{1/2} (4\bar{5})^{8}_{1} \rangle_{1/2} \nonumber \\
  &-| [(12)^{\bar{3}}_{1}3]^{8}_{1/2} (4\bar{5})^{8}_{1} \rangle_{1/2}\big), \nonumber \\
  \frac{1}{\sqrt{2}}\left(\phi_{1}\chi_{10}-\phi_{2}\chi_{8}\right)=&\frac{1}{\sqrt{2}}\big(| [(12)^{6}_{0}3]^{8}_{1/2} (4\bar{5})^{8}_{0} \rangle_{1/2} \nonumber \\
  &-| [(12)^{\bar{3}}_{1}3]^{8}_{1/2} (4\bar{5})^{8}_{0} \rangle_{1/2}\big), \nonumber \\
  \phi_{3}\chi_{6} =& | [(12)^{\bar{3}}_{1}3]^{1}_{3/2} (4\bar{5})^{1}_{1} \rangle_{1/2}.
\end{align}

The wave function of pentaquark $QQnQ^{\prime}\bar{n}$.

a.$J^{P}=5/2^{-}:$
\begin{align}
  \phi_{2}\chi_{1} = | [(12)^{\bar{3}}_{1}3]^{8}_{3/2} (4\bar{5})^{8}_{1} \rangle_{5/2}, \nonumber \\
  \phi_{3}\chi_{1} = | [(12)^{\bar{3}}_{1}3]^{1}_{3/2} (4\bar{5})^{1}_{1} \rangle_{5/2}.
\end{align}

b.$J^{P}=3/2^{-}:$
\begin{align}
  \phi_{1}\chi_{5} = | [(12)^{6}_{0}3]^{8}_{1/2} (4\bar{5})^{8}_{1} \rangle_{3/2}, \nonumber \\
  \phi_{2}\chi_{2} = | [(12)^{\bar{3}}_{1}3]^{8}_{3/2} (4\bar{5})^{8}_{1} \rangle_{3/2}, \nonumber \\
  \phi_{2}\chi_{3} = | [(12)^{\bar{3}}_{1}3]^{8}_{3/2} (4\bar{5})^{8}_{0} \rangle_{3/2}, \nonumber \\
  \phi_{2}\chi_{4} = | [(12)^{\bar{3}}_{1}3]^{8}_{1/2} (4\bar{5})^{8}_{1} \rangle_{3/2}, \nonumber \\
  \phi_{3}\chi_{2} = | [(12)^{\bar{3}}_{1}3]^{1}_{3/2} (4\bar{5})^{1}_{1} \rangle_{3/2}, \nonumber \\
  \phi_{3}\chi_{3} = | [(12)^{\bar{3}}_{1}3]^{1}_{3/2} (4\bar{5})^{1}_{0} \rangle_{3/2}, \nonumber \\
  \phi_{3}\chi_{4} = | [(12)^{\bar{3}}_{1}3]^{1}_{1/2} (4\bar{5})^{1}_{1} \rangle_{3/2}.
\end{align}

c.$J^{P}=1/2^{-}:$
\begin{align}
  &\phi_{1}\chi_{9} \ = | [(12)^{6}_{0}3]^{8}_{1/2} (4\bar{5})^{8}_{1} \rangle_{1/2}, \nonumber \\
  &\phi_{1}\chi_{10} = | [(12)^{6}_{0}3]^{8}_{1/2} (4\bar{5})^{8}_{0} \rangle_{1/2}, \nonumber \\
  &\phi_{2}\chi_{6} \ = | [(12)^{\bar{3}}_{1}3]^{8}_{3/2} (4\bar{5})^{8}_{1} \rangle_{1/2}, \nonumber \\
  &\phi_{2}\chi_{7} \ = | [(12)^{\bar{3}}_{1}3]^{8}_{1/2} (4\bar{5})^{8}_{1} \rangle_{1/2}, \nonumber \\
  &\phi_{2}\chi_{8} \ = | [(12)^{\bar{3}}_{1}3]^{8}_{1/2} (4\bar{5})^{8}_{0} \rangle_{1/2}, \nonumber \\
  &\phi_{3}\chi_{6} \ = | [(12)^{\bar{3}}_{1}3]^{1}_{3/2} (4\bar{5})^{1}_{1} \rangle_{1/2}, \nonumber \\
  &\phi_{3}\chi_{7} \ = | [(12)^{\bar{3}}_{1}3]^{1}_{1/2} (4\bar{5})^{1}_{1} \rangle_{1/2}, \nonumber \\
  &\phi_{3}\chi_{8} \ = | [(12)^{\bar{3}}_{1}3]^{1}_{1/2} (4\bar{5})^{1}_{0} \rangle_{1/2}.
\end{align}

The wave function of pentaquark $QQ^{\prime}nQ\bar{n}$.

a.$J^{P}=5/2^{-}:$
\begin{align}
  \phi_{1}\chi_{1}=|[(12)^{6}_{1}3]^{8}_{3/2} (4\bar{5})^{8}_{1}\rangle_{5/2}, \nonumber \\
  \phi_{2}\chi_{1}=|[(12)^{\bar{3}}_{1}3]^{8}_{3/2} (4\bar{5})^{8}_{1}\rangle_{5/2}, \nonumber \\
  \phi_{3}\chi_{1}=|[(12)^{\bar{3}}_{1}3]^{1}_{3/2} (4\bar{5})^{1}_{1}\rangle_{5/2}.
\end{align}

b.$J^{P}=3/2^{-}:$
\begin{align}
  \phi_{1}\chi_{2}=|[(12)^{6}_{1}3]^{8}_{3/2} (4\bar{5})^{8}_{1}\rangle_{3/2}, \nonumber \\
  \phi_{1}\chi_{3}=|[(12)^{6}_{1}3]^{8}_{3/2} (4\bar{5})^{8}_{0}\rangle_{3/2}, \nonumber \\
  \phi_{1}\chi_{4}=|[(12)^{6}_{1}3]^{8}_{1/2} (4\bar{5})^{8}_{1}\rangle_{3/2}, \nonumber \\
  \phi_{1}\chi_{5}=|[(12)^{6}_{0}3]^{8}_{1/2} (4\bar{5})^{8}_{1}\rangle_{3/2}, \nonumber \\
  \phi_{2}\chi_{2}=|[(12)^{\bar{3}}_{1}3]^{8}_{3/2} (4\bar{5})^{8}_{1}\rangle_{3/2}, \nonumber \\
  \phi_{2}\chi_{3}=|[(12)^{\bar{3}}_{1}3]^{8}_{3/2} (4\bar{5})^{8}_{0}\rangle_{3/2}, \nonumber \\
  \phi_{2}\chi_{4}=|[(12)^{\bar{3}}_{1}3]^{8}_{1/2} (4\bar{5})^{8}_{1}\rangle_{3/2}, \nonumber \\
  \phi_{2}\chi_{5} = | [(12)^{\bar{3}}_{0}3]^{8}_{1/2} (4\bar{5})^{8}_{1} \rangle_{3/2}, \nonumber \\
  \phi_{3}\chi_{2} = | [(12)^{\bar{3}}_{1}3]^{1}_{3/2} (4\bar{5})^{1}_{1} \rangle_{3/2}, \nonumber \\
  \phi_{3}\chi_{3} = | [(12)^{\bar{3}}_{1}3]^{1}_{3/2} (4\bar{5})^{1}_{0} \rangle_{3/2}, \nonumber \\
  \phi_{3}\chi_{4} = | [(12)^{\bar{3}}_{1}3]^{1}_{1/2} (4\bar{5})^{1}_{1} \rangle_{3/2}, \nonumber \\
  \phi_{3}\chi_{5} = | [(12)^{\bar{3}}_{0}3]^{1}_{1/2} (4\bar{5})^{1}_{1} \rangle_{3/2}.
\end{align}

c.$J^{P}=1/2^{-}:$
\begin{align}
  &\phi_{1}\chi_{6} = | [(12)^{6}_{1}3]^{8}_{3/2} (4\bar{5})^{8}_{1} \rangle_{1/2}, \nonumber \\
  &\phi_{1}\chi_{7} = | [(12)^{6}_{1}3]^{8}_{1/2} (4\bar{5})^{8}_{1} \rangle_{1/2}, \nonumber \\
  &\phi_{1}\chi_{8} = | [(12)^{6}_{1}3]^{8}_{1/2} (4\bar{5})^{8}_{0} \rangle_{1/2}, \nonumber \\
  &\phi_{1}\chi_{9} = | [(12)^{6}_{0}3]^{8}_{1/2} (4\bar{5})^{8}_{0} \rangle_{1/2}, \nonumber \\
  &\phi_{1}\chi_{10} = | [(12)^{6}_{0}3]^{8}_{1/2} (4\bar{5})^{8}_{0} \rangle_{1/2}, \nonumber \\
  &\phi_{2}\chi_{6}  = | [(12)^{\bar{3}}_{1}3]^{8}_{3/2} (4\bar{5})^{8}_{1} \rangle_{1/2}, \nonumber \\
  &\phi_{2}\chi_{7}  = | [(12)^{\bar{3}}_{1}3]^{8}_{1/2} (4\bar{5})^{8}_{1} \rangle_{1/2}, \nonumber \\
  &\phi_{2}\chi_{8}  = | [(12)^{\bar{3}}_{1}3]^{8}_{1/2} (4\bar{5})^{8}_{0} \rangle_{1/2}, \nonumber \\
  &\phi_{2}\chi_{9}  = | [(12)^{\bar{3}}_{0}3]^{8}_{1/2} (4\bar{5})^{8}_{1} \rangle_{1/2}, \nonumber \\
  &\phi_{2}\chi_{10} = | [(12)^{\bar{3}}_{0}3]^{8}_{1/2} (4\bar{5})^{8}_{0} \rangle_{1/2}, \nonumber \\
  &\phi_{3}\chi_{6}  = | [(12)^{\bar{3}}_{1}3]^{1}_{3/2} (4\bar{5})^{1}_{1} \rangle_{1/2}, \nonumber \\
  &\phi_{3}\chi_{7}  = | [(12)^{\bar{3}}_{1}3]^{1}_{1/2} (4\bar{5})^{1}_{1} \rangle_{1/2}, \nonumber \\
  &\phi_{3}\chi_{8} = | [(12)^{\bar{3}}_{1}3]^{1}_{1/2} (4\bar{5})^{1}_{0} \rangle_{1/2}, \nonumber \\
  &\phi_{3}\chi_{9}  = | [(12)^{\bar{3}}_{0}3]^{1}_{1/2} (4\bar{5})^{1}_{1} \rangle_{1/2}, \nonumber \\
  &\phi_{3}\chi_{10} = | [(12)^{\bar{3}}_{0}3]^{1}_{1/2} (4\bar{5})^{1}_{0} \rangle_{1/2}.
\end{align}

\end{document}